\documentclass{elsart1p}
\usepackage{graphicx}
\usepackage{amssymb}

\newcommand{\Kzs}{\mbox{$\mathrm {K^0_S}$}}

\newcommand{\Jpsi} {\mbox{J\kern-0.05em /\kern-0.05em$\psi$}}

\begin{document}
\begin{frontmatter}
%
%
%
%
%
\title{First Results from the ALICE experiment at the LHC}
%
%

\author{J. Schukraft}  

For the ALICE Collaboration

\address{CERN, Geneva, Switzerland}

\begin{abstract}
 After close to 20 years of preparation, the dedicated heavy ion experiment ALICE took first data with proton collisions at the LHC starting in November 2009 and first Pb--Pb data in November 2010. This article summarizes initial operation and performance of ALICE as well as first results from both pp and 
Pb--Pb collisions.

\end{abstract}

\end{frontmatter}

\section{Introduction}
ALICE (A Large Ion Collider Experiment) is the dedicated heavy ion experiment at the CERN LHC, optimised to study matter under extreme conditions of temperature and pressure -- the Quark-Gluon Plasma (QGP) -- in collisions between heavy nuclei. With an energy up to 30 times higher than RHIC, the current energy frontier machine for heavy ion collisions at BNL, we expect both a very different type of QGP, e.g. in terms of initial temperature, lifetime and system volume, and an abundance of hard signals like jets and heavy quarks which serve as probes to study QGP properties. Data taking with pp (and later p-nucleus) is equally important, primarily to collect comparison data for the heavy ion program.  In addition, a number of measurements concerning soft and semi-hard QCD processes in Minimum Bias (MB) and high multiplicity pp collisions are part of the initial physics program~\cite{Carminati:2004fp}, taking advantage of the specific and complementary capabilities of the detector. The large MB sample will also provide a detailed characterisation of global event properties over a range of LHC energies, which will be very useful for tuning Monte Carlo generators to better describe the QCD background underlying searches for new physics. With the high multiplicity pp sample, which is enriched with a dedicated trigger to reach particle densities comparable to nuclear collisions at lower energies, we can test if heavy ion like features become apparent as multiplicity increases or if they are dominated by hard multijet final states.

ALICE consists of a central part, which measures hadrons, electrons and photons, and a forward spectrometer to measure muons. The central part, which covers polar angles from $45^0$ to $135^0$ over the full azimuth, is embedded in the large L3 solenoidal magnet. It consists of an inner tracking system (ITS) of high-resolution silicon detectors, a cylindrical TPC, three particle identification arrays of Time-of-Flight (TOF), Cerenkov (HMPID) and Transition Radiation (TRD) counters, and two single-arm electromagnetic calorimeters (high resolution PHOS and large acceptance EMCAL). The forward muon arm ($2^0-9^0$) consists of a complex arrangement of absorbers, a large dipole magnet, and 14 stations of tracking and triggering chambers. Several smaller detectors for triggering and multiplicity measurements (ZDC, PMD, FMD, T0, V0) are located at small angles. 
The main design features include a robust and redundant tracking over a limited region of pseudorapidity, designed to cope with the very high particle density of nuclear collisions, a minimum of material in the sensitive tracking volume (10\% radiation length between vertex and outer radius of the TPC) to reduce multiple scattering, and several detector systems dedicated to particle identification over a large range in momentum~\cite{JSPLHC2010,Evans:2009zz}.

The layout of the ALICE detector and its eighteen different subsystems are described in detail in~\cite{ALICEdet}. The experiment is fully installed, commissioned and operational, with the exception of two systems (TRD and EMCAL) which were added more recently and are only now nearing the end of construction. Both systems have currently about 40\% of their active area installed and will be completed during the winter shutdown in 2010/11 (EMCAL) and 2012/13 (TRD). 

The LHC started operating in November 2009 with pp collisions at $\sqrt{s} = 900$ GeV and reached its current maximum energy of 7 TeV in March 2010. The primary ALICE goal for pp was to collect about $10^{9}$  minimum bias collisions under clean experimental conditions.
Therefore the experiment was running for much of 2010 in a special low luminosity mode, in which the LHC beams are separated by $ 3-5 \sigma$ in the ALICE interaction region, to keep event pile-up below a few percent per bunch crossing. The total pp data sample collected at 7 TeV (900 GeV) corresponds to some 700 M (7 M) MB triggers, 50 M muon triggers and about 20 M high multiplicity events. In addition, a very short test run was taken at 2.36 TeV with only a small subset of detectors being read out.

First heavy ion collisions at 2.76 TeV/nucleon started just prior to this conference in November 2010, after only a few days of switching over from the pp set-up. Like for pp, the LHC worked exceedingly well also for heavy ions, with a steep increase in luminosity, reaching about $2~10^{25} \rm cm^{-2} \rm s^{-1}$ towards the end of the run and yielding some 30 M nuclear MB interactions on tape. 

The many years of preparation, analysis tuning with simulations, and detector commissioning with cosmics during much of 2008/9 paid off with most of the detector components working 'right out of the box' and rather close to performance specifications, with both pp and Pb--Pb collisions . Within days all experiments could show first qualitative (and some quantitative) results. By the time of writing these proceedings, many of the results presented here have been already submitted for publication; they are therefore only briefly summarised in these proceedings (see also~\cite{Schukraft:2010ru}). Further details and references can be found in the original publications and in the large number of ALICE contributions presented at this conference. 

\section{Results from pp collisions}
The charged particle density ${\rm d}N_{\rm ch}/{\rm d}\eta$ as well as the multiplicity distributions were measured at 0.9, 2.36 and 7 TeV for  inelastic and non-single diffractive collisions~\cite{:2009dt,Aamodt:2010pp,Aamodt:2010ft}. The energy dependence of the multiplicity is well described by a power law in energy, $s^{0.1}$, and increases significantly more strongly than predicted by most event generators. Most of this stronger increase happens in the tail of the multiplicity distribution, i.e. for events with much larger than average multiplicity. Likewise, neither the transverse momentum distribution at 900 GeV nor the dependence of average $p_T$ on $N_{\rm ch}$ is well described by various versions of generators~\cite{Aamodt:2010my}, in particular when including low momentum particles ($p_T < 500$ MeV). The shape of the $p_T$ spectrum as a function of multiplicity hardly changes below  0.8 GeV (which includes the large majority of all particles), whereas the power law tail increases rapidly with event multiplicity above about 1-2 GeV.

Bose Einstein (HBT) correlations between identical particles are used to measure the space-time evolution of the dense matter system created in heavy ion collisions; their interpretation in elementary reactions ($e^+e^-, pp$) is controversial. However, HBT measurements at LHC are important also in pp as a comparison for the heavy ion data, because non-HBT correlations, which must be subtracted, are expected to increase with energy (e.g. via increased jet and mini-jet activity in MB events). They should also clarify if systematic trends, seen e.g. as a function of multiplicity and pair momentum $k_T$, differ between pp and heavy ions. The experimental trends observed at both 900 GeV and 7 TeV~\cite{Aamodt:2010jj, Aamodt:2011kd} can be summarised as follows:
HBT radii, which are typically of order 1 fm in elementary collisions, increase smoothly with multiplicity as previously observed both at low (ISR, RHIC) and at high (FNAL) energies,  a trend which is qualitatively similar to the one seen in heavy ion collisions. 
However, the decrease of the radius with pair momentum appears to be much weaker in our MB data than seen at the Tevatron; it becomes increasingly significant only for high pp multiplicities.
However, the decrease of radius with pair momentum seems to be much weaker in our MB data than the one seen at the Tevatron by E735; it becomes increasingly significant only for high pp multiplicities. We observe, both in the data (with unlike-sign pairs) as well as in the Monte Carlo (both Pythia and Phojet), strong particle correlations at small relative pair momentum, i.e. in the region of the HBT effect, which had to be carefully studied and subtracted. If we evaluate the HBT radii without subtracting non-HBT correlations, as has been done traditionally in the past, we observe a much stronger momentum dependent decrease of the radius, as these minijet-like correlations become more pronounced at higher momentum, effectively simulating a decreasing radius. A detailed comparison in the various HBT components (side/long/out) between pp and nuclear reactions as a function of ${\rm d}N_{\rm ch}/{\rm d}\eta$ and $k_T$ reveals qualitatively similar, but quantitatively distinct trends. The question if high multiplicity pp collisions show 'collectivity' or 'expansion-like' features requires further digestion of these results and comparison with models.

At the LHC, by far the highest energy proton--proton collider, we have studied baryon transport over very large rapidity intervals 
by measuring the antiproton-to-proton ratio at mid-rapidity~\cite{Aamodt:2010dx} in order to discriminate between various theoretical models of baryon stopping.
Baryon number transport over large gaps in rapidity is often described in terms of a nonlinear three gluon configuration called 'baryon string junction'. The dependence of this process on the size of the rapidity gap has been a longstanding issue (for large gaps, where it should be dominant), with advocates for both very weak and rather strong dependencies.
In either case, the $\bar{p}/p$ ratio at LHC is very close to 1.0, with the difference between various models only of the order of a few percent. So this ratio must be measured with high precision. While in the ratio many instrumental effects cancel, the very large difference between $p$ and $\bar{p}$ cross section for both elastic (track can get lost) and inelastic (particle can be absorbed)  reactions with the detector material lead to corrections of order 10\%, even in the very thin central part of ALICE. As the corrections are much larger than the effect, a very precise knowledge of the detector material as well as of the relevant cross section values at low momentum is required. The former was measured with the data via photon conversions with a relative precision of $< 7\%$ (absolute precision of better than 7 per mil radiation length!); the latter had to be cross checked with experimental data and the Fluka transport code since all available versions of Geant overestimate the $\bar{p}$ cross sections by up to a factor of five. The $\bar{p}/p$ ratio was found to be compatible with 1.0 at 7 TeV and 4\% below 1.0 at 900 GeV, with an experimental uncertainty of about 1.5\%, dominated by the systematic error. This result favours models which predict a strong suppression of baryon transport over large gaps; they agree very well with standard event generators but not with those that have implemented enhanced proton stopping.

The yields and $p_T$ spectra of identified stable charged ($\pi$, K, p) and neutral strange particles 
($\Kzs,\Lambda, \Xi, \phi$) have been measured for the small 900 GeV data set taken in 2009~\cite{Aamodt:2011zj,Collaboration:2010vf}. In most cases Phojet as well as several Pythia tunes are well below the data -- by factors of two to almost five -- and more so at high $p_T$ and for the heavier particles ($\Lambda, \Xi$). The ratio of $\Lambda$ to $\Kzs$ agrees very well with the STAR data at 200 GeV but is significantly below the ratio measured by UA1 and CDF. This discrepancy merits further investigation; it could be due to differences between the experiments in the acceptance, triggers, or correction for feed-down from weak decays. Baryon to meson ratios are of particular interest, as they rise well above unity in nuclear collisions at RHIC. This 'meson-baryon' anomaly has been interpreted in the context of coalescence models as an indirect sign of the QGP, in which case it would not be obvious why similar ratios should be reached already in minimum bias pp interactions. 

\section{Results from Pb--Pb collisions}
The main aim of ultra-relativistic heavy ion physics is to {\em search} for the QGP, to {\em measure} its properties, and along the way {\em discover} new aspects of QCD in the regime where the strong interaction is strong indeed. With the fantastic results from RHIC, I consider the {\em search} for the QGP essentially over and the {\em discovery} phase well on the way. However, {\em precision measurements} of QGP parameters are just starting (precision is meant here in the context of non-pertubative QCD, where a factor of two is already respectable). In this program, the LHC has some unique advantages and is complementary to RHIC in other aspects. We expect significant differences  at the higher energy in terms of energy density, lifetime and volume of the state of matter created in the collisions, and rare 'hard probes' (jets, heavy flavours, quarkonia) will be abundantly produced. In this new environment, we can test and validate the 'heavy ion standard model' (HI-SM), which describes the QGP as a strongly interacting, very opaque, almost perfect fluid. This model has emerged over the last 10 years from RHIC, and it would be a non trivial test to see how it fares at much higher energy. Once we have verified that the global event characteristics (e.g. energy density, volume, lifetime) of matter at LHC are indeed rather different, but that evolution and intrinsic properties are still well described  by the standard model, we can embark on the program of precision measurements of the QGP parameters (e.g. viscosity, equation-of-state, transport coefficients, Debye screening mass,..). And along the way, we may well discover some more surprises...

\begin{figure}[!t]
\centerline{\includegraphics[width=0.9\textwidth]{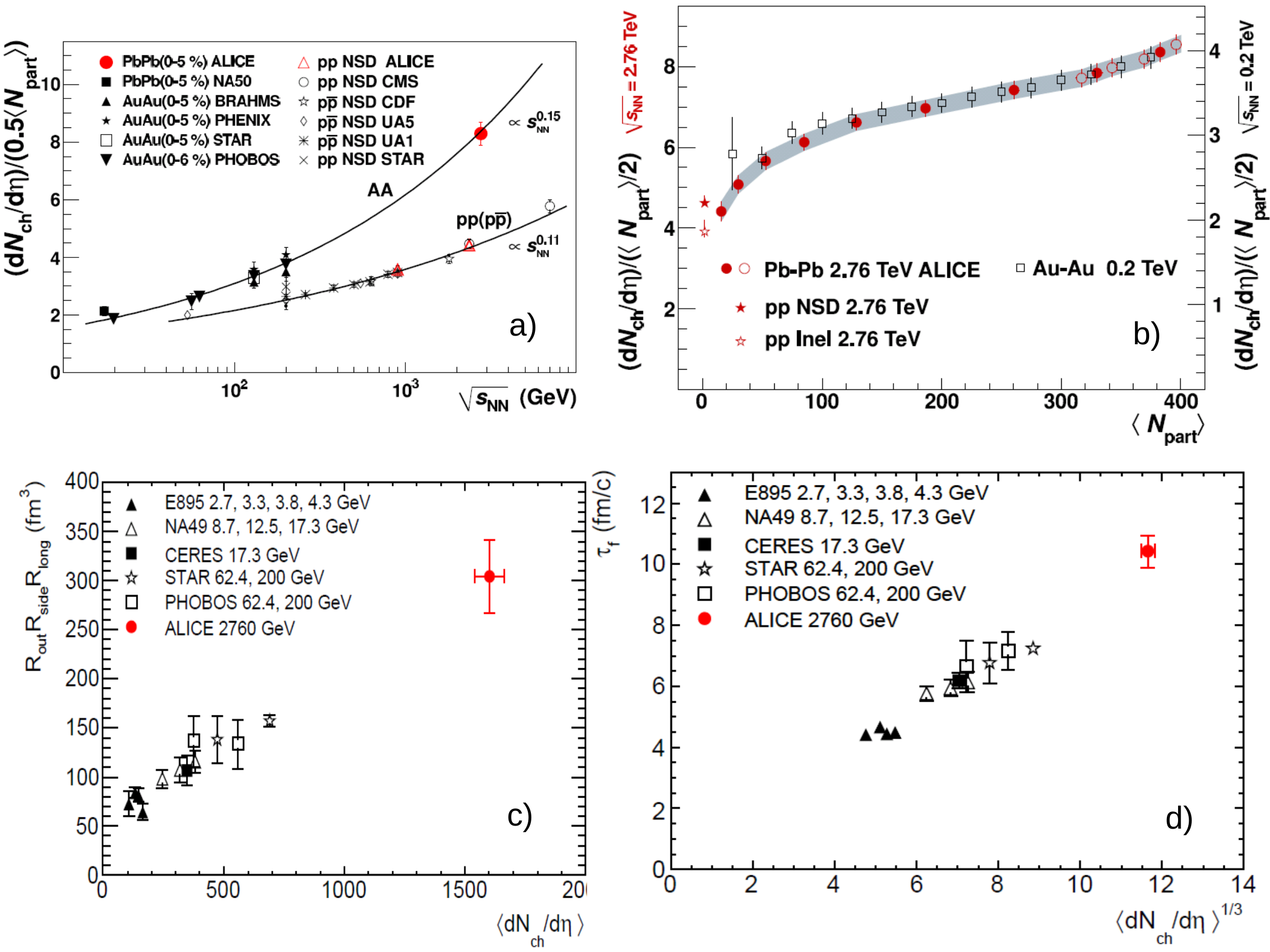}}

\caption{Top left: Charged particle pseudorapidity density per participant versus centre of mass energy for pp and AA collisions. Top right: Charged particle density per participant versus number of participants $N_{\rm part}$ for LHC and RHIC data. Bottom left: Product of the three pion HBT radii at $k_T$ = 0.3 GeV/c compared to central gold and lead collisions at lower energies. Bottom right: The decoupling time $\tau_{f}$ compared to results from lower energies. 
Figures taken from refs~\cite{Aamodt:2010pb, Collaboration:2010cz, Aamodt:2011mr}.}
\label{fig1}
\end{figure}

The first, and most anticipated, results concerned global event characteristics. When ALICE was conceived in the early '90s, the charged particle density predicted for central Pb--Pb collisions at LHC was extremely uncertain, varying between 2000 and $>4000$ charge particles per unit rapidity. With RHIC results, the uncertainties came down quite a bit, as did the central value, with most predictions clustering in the range 1000 - 1700~\cite{Abreu:2007kv}.
The value finally measured at LHC, ${\rm d}N_{\rm ch}/{\rm d}\eta \approx 1600$~\cite{Aamodt:2010pb}, was right in this range, if somewhat on the high side. 
From the measured multiplicity in central collisions one can derive a rough estimate of the energy density, which gives at least a factor 3 above RHIC. The corresponding increase of the initial temperature is about 30\%, even with the conservative assumption that the formation time $\tau_0$ does not decrease from RHIC to LHC.
When combined with lower energy data, the charged particle production per participant, $N_{\rm part}$, rises stronger with energy than in pp, approximately with $s^{0.15}$ (Figure~\ref{fig1} top left). Even more surprising is the fact that the centrality dependence of ${\rm d}N_{\rm ch}/{\rm d}\eta$~\cite{Collaboration:2010cz} (Figure~\ref{fig1} top right)
is practically identical to Au--Au at RHIC (at least for $N_{\rm part} >50$), despite the fact that impact parameter dependent shadowing/saturation would be expected to be much stronger at LHC (much smaller Feynman-x). And indeed, models with strong shadowing (like the latest version of Hijing) and different saturation-type calculations do describe the impact parameter dependence best. 
Does the similarity between the two energies then mean that saturation saturates already at RHIC? Or is it just that the nuclear geometry reigns supreme?�

\begin{figure}[!t]
\centerline{\includegraphics[width=0.90\textwidth]{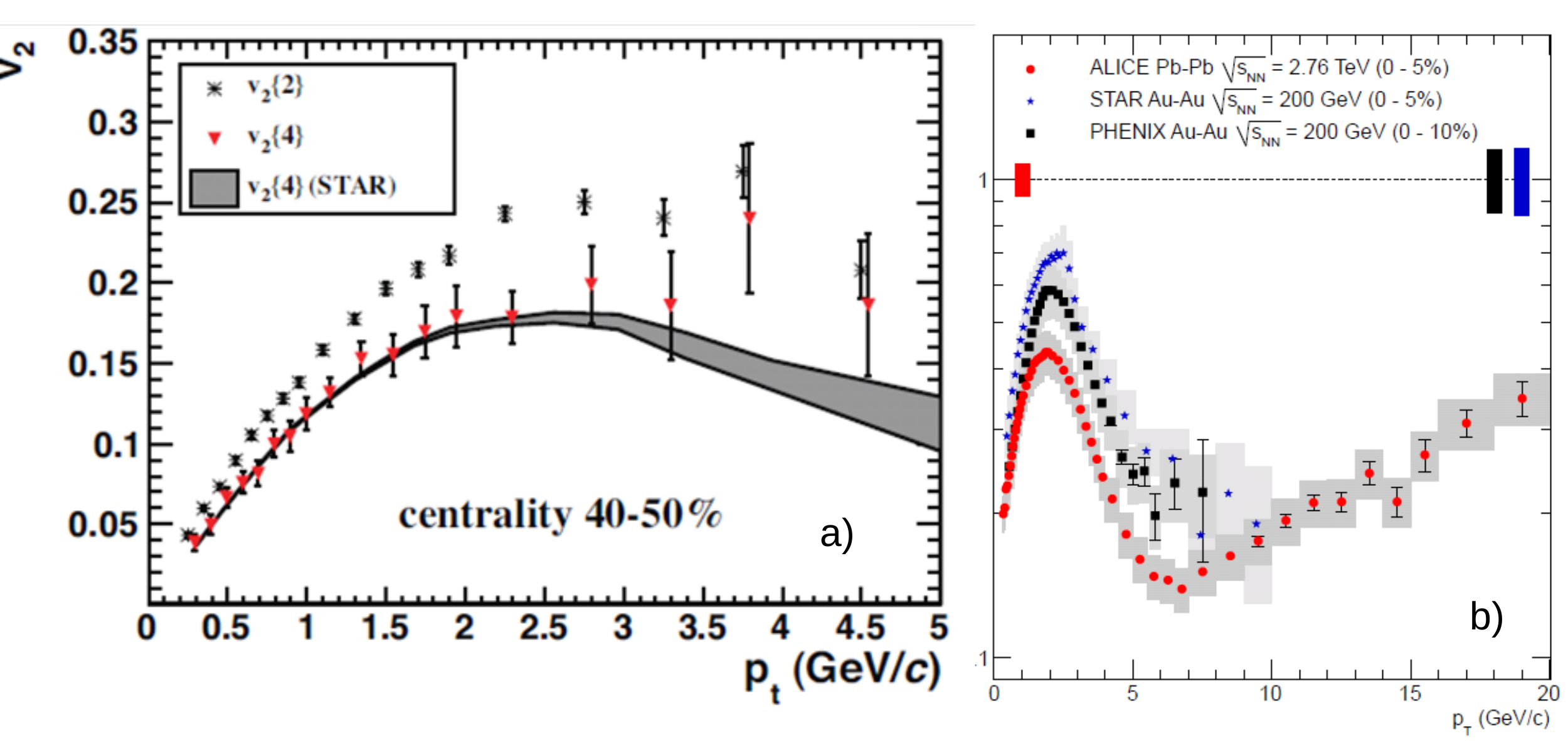}}

\caption{Left: Elliptic flow coefficient $v2$ versus $p_T$ compared to RHIC data. Right: Nuclear modification factor $R_{AA}$ for central Pb--Pb at LHC compared to RHIC data. Figures taken from refs~\cite{Aamodt:2010pa, Aamodt:2010jd}.}
\label{fig2}
\end{figure}

The freeze-out volume and total lifetime of the created system was measured with identical particle interferometry (HBT)~\cite{Aamodt:2011mr}. Compared to top RHIC energy, the 'volume of homogeneity' ((Figure~\ref{fig1} bottom left) increases by a factor 2 (to 300 $fm^3$) and the system lifetime (Figure~\ref{fig1} bottom right) increases by about 20\% (to 10 fm/c), pretty much in line with the predictions from hydrodynamics. Also a more detailed study of the out/long/side radii as a function of pair momentum shows very good agreement with hydro models, at least with those who do describe well also the RHIC data. After years of confusion and discrepancy with data (the dark ages of the 'HBT puzzle'), interferometry is again a predictive and consistent part of the HI-SM.

The most critical test of the heavy ion standard model however comes from the measurement of the elliptic flow at LHC, the pillar which supports the 'fluid' interpretation of the QGP (perfect or otherwise). Assuming only small or no changes in the fluid properties (EoS, viscosity) between RHIC and LHC,  hydro predicts firmly that the elliptic flow coefficient $v2$, measured as a function of $p_T$, should not depend on beam energy. As shown in Figure~\ref{fig2} (left), this prediction was confirmed very quickly and with good precision~\cite{Aamodt:2010pa}. The $p_T$ integrated flow values however do increase compared to RHIC by some 30\% for mid-central collisions, because the average $p_T$ is significantly higher at LHC. While $<p_T>$ increases also in pp with energy, because hard and semi-hard processes become more important, hydro predicts in addition an increase in the radial flow velocity leading to a characteristic $p_T$ and mass dependence of the spectra. It will be very interesting to see, once identified particle spectra will be available, if also this prediction is borne out (as everyone assumes it is), and if the radial flow regime extends to even higher momentum than at RHIC (here predictions are less firm).

With the heavy ion standard model having passed its first tests (HBT, $v2$) with flying colours, the program of precision measurements is now starting at LHC. For example, a major advance in quantifying the shear viscosity, which at RHIC was measured to be within a factor 2-4 of the quantum limit of a perfect fluid, will require a better estimate of remaining non-flow contributions, as well as a better constraint on the initial conditions, i.e. the geometry of the collision zone (and its fluctuations) which drive the various flow components. Progress is already being made on both fronts, in particular by looking in more detail at the centrality dependence and  at higher Fourier components ($v2, v3, v4,..$).

\begin{figure}[!t]
\centerline{\includegraphics[width=0.9\textwidth]{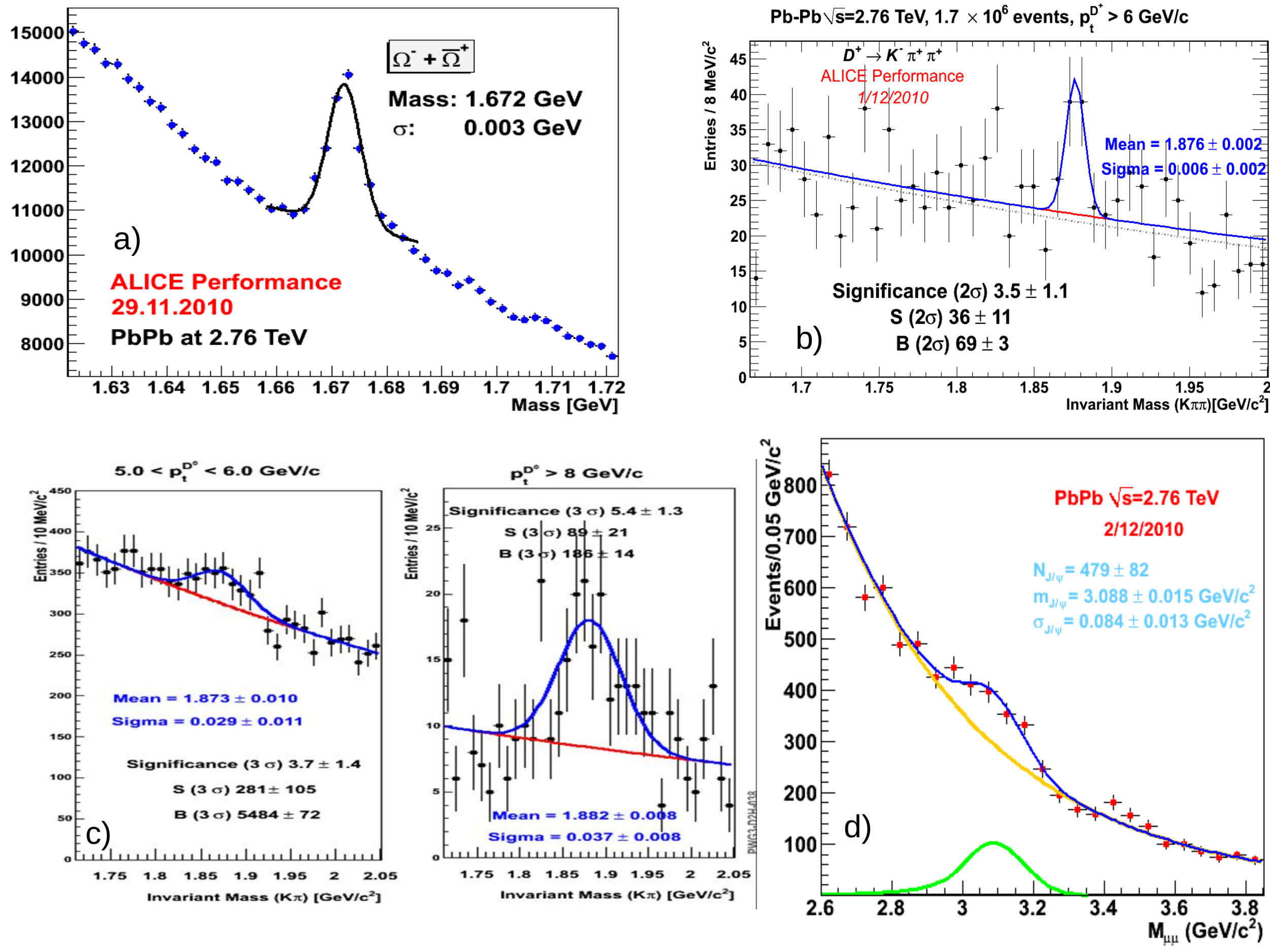}}

\caption{ Top left: Reconstructed $\Omega$ decays. Top right and bottom left: Reconstructed hadronic charm decays. Bottom right: Dimuon invariant mass distribution around the $\Jpsi$ mass. }
\label{fig3}
\end{figure}

The energy advantage of LHC is most evident in the area of parton energy loss and jet quenching, where the kinematic reach vastly exceeds the one available at RHIC. With high $p_T$ jets easily sticking out of the soft background, jet quenching is qualitatively evident already by visual inspection of charged jets in the TPC where a striking jet energy imbalance develops for central collisions. A quantitative analysis will however take more time, in order to see how much energy is actually lost (i.e. measure the transport coefficient), how and where it is lost (multiple soft versus few hard scatterings), and if there is a response of the medium to this local energy deposit (shock waves, mach cones). Eventually, an answer to all these questions will require measuring the medium modification of the fragmentation function, down to low ($<$ 5 GeV) or even very low ($<$ 2 GeV) transverse momentum. In the meantime, we have measured the nuclear modification factor $R_{AA}$ of inclusive charged particle momentum distributions out to 20 GeV~\cite{Aamodt:2010jd}, where the spectra are dominated by leading jet fragments. As shown in Figure~\ref{fig2} (right), the $R_{AA}$ ratio has a minimum at around 6 GeV, where the suppression is modestly stronger than at RHIC, but then rises again smoothly towards higher momentum. This latter feature is not evident in published RHIC data, and while such a rise was qualitatively predicted by some models for LHC~\cite{Abreu:2007kv}, it looks stronger at first sight. However, initial state effects (shadowing/saturation), which presumably are very strong at LHC and which should depend on both impact parameter and momentum transfer, can complicate a straight forward interpretation of the data and the comparison between different beam energies. It will be interesting to see how this result will fit into the overall picture of jet quenching.

Finally, as an appetizer of results to come, Figure~\ref{fig3} shows some first performance plots on strange particle and heavy flavour production in Pb--Pb for a fraction of the collected statistics (for details and additional results see the other contributions of ALICE to this conference): the $\Omega$ (top left) and fully reconstructed hadronic charm decays in the central barrel region (top right, bottom left) and the $\Jpsi$ decaying into muons in the forward muon arm (bottom right). However, as in the case of $R_{AA}$, charm and quarkonia production and their distributions may be strongly modified by initial state effects which could mask the medium modification which is the prime interest. Therefore, like at RHIC, a pA comparison run at LHC may be required sooner rather than later.

\section{Summary}
After two decades of design, R\&D, construction, installation, commissioning and simulations, the ALICE experiment has 'hit the ground running' since LHC started its operation at the end of 2009. Most systems are fast approaching design performance, and physics analysis is well underway. With the first heavy ion collisions a mere four weeks ago, a new era has started for ultra-relativistic heavy ion physics and we can look forward to at least a decade of exciting (and hopefully revealing) new results.

\end{document}